\documentclass[prl,floatfix,twocolumn,showpacs]{revtex4}
\usepackage{amsmath}
\usepackage{amsfonts}
\usepackage{mathrsfs}
\usepackage{epsfig}
\usepackage{graphicx}
\usepackage{dcolumn}
\usepackage{amsmath}
\usepackage{url}

\begin{document}

\title{Dynamical order, disorder and propagating defects in
%Propagating phase defects, Spatio-temporal patterns in 
homogeneous system of relaxation oscillators}
%\title{Anti-phase oscillations and Turing patterns in a lattice of
%coupled relaxation oscillators}
\author{Rajeev Singh and Sitabhra Sinha}
\affiliation{The Institute of Mathematical Sciences, CIT Campus,
Taramani, Chennai 600113, India.}
\date{\today}
\begin{abstract}
Reaction-diffusion (RD) mechanisms in chemical and biological
systems can yield a variety of patterns that may
be functionally important.
%such as in morphogenesis.
%Motivated by recent experiments on patterns generated 
%in oscillatory reactive media, 
We show that diffusive coupling through the inactivating component
in a generic model of coupled relaxation oscillators give rise to a
wide range of spatio-temporal phenomena. Apart from analytically
explaining the genesis of anti-phase
synchronization and spatially patterned oscillatory death regimes in
the model system,
%which have been seen experimentally, 
we report the existence of a chimera
state, characterized by spatial co-occurrence of patches
with distinct dynamics.
%in a homogeneous medium.
We also observe propagating phase defects in both one- and two-dimensional
media resembling persistent structures in cellular
automata, whose interactions may be used for computation in RD systems.
%abstract should be 600 characters including space
\end{abstract}

\maketitle

%%%%
%\newpage
%%%%
%\section*{Introduction}
Understanding how patterns develop in chemical and biological systems, 
e.g., in the course of morphogenesis~\cite{Morelli12}, is one of the enduring challenges
of modern science~\cite{Cross1993}. Investigating pattern formation in systems of coupled
biochemical oscillators is a promising approach towards this objective
and has been applied to study, for instance, the
temporal organization of gene activity in cells during the course of
development~\cite{Goodwin64}. The patterns seen in such systems are the 
dynamical attractors
that can change with the underlying
system parameters as the environment evolves with time. 
Such phenomena have been sought to be explained by reaction-diffusion
models that exhibit patterns such as stripes and spots under
specific conditions, viz., when an inhibitory chemical species diffuses
faster than an activating species, by destabilizing the
homogeneous steady state~\cite{Kondo10}. Indeed, generalizations of
such processes involving
differential excitatory and inhibitory interactions between elements
is seen in a wide variety of complex systems
\cite{Meinhardt72,Gong12,Butler12}. Recently such mechanisms have also
been proposed as a possible basis for computation in biological and
chemical systems~\cite{Liang1995,Epstein2007}.

Time-invariant patterns that are seen in the models mentioned
above are, however, only a small segment of the variety seen in nature,
many of which exhibit periodic activity. Thus, extending the concept
of reaction-diffusion mechanism to systems of interacting relaxation 
oscillators may allow investigation of spatio-temporal
patterns in biological systems, where oscillations are
observed across many spatial and temporal scales, e.g., from the periodic
variations of intracellular molecular
concentrations~\cite{Orchard83} to changes in the activity levels of
different brain
areas~\cite{Buzsaki04}.
%and from genetic expression levels that 
%vary over the day~\cite{Vitaterna94} to menstrual cycles~\cite{Stern98}.
Coherent dynamics of such oscillators can produce functionally
important collective behavior such as synchronization~\cite{Kiss02} giving
rise to different biological rhythms~\cite{Singh12}. However, synchronized
oscillations is only one of a number of possible collective phenomena
that can emerge from such interactions. For example, a recent set of
experiments on coupled chemical oscillators in a microfluidic
device~\cite{Toiya08}
have shown that
anti-phase synchronization as well as spatially heterogeneous
oscillator death
states~\cite{BarEli85} % reference for oscillator death
can occur under different conditions.
Extending the mechanism
of coupling by lateral inhibition (e.g., 
via a rapidly diffusing inhibitory chemical species) to
%elements at equilibrium state.
arrays of coupled relaxation oscillators, used for modeling
biological periodic activity, may reveal the underlying mechanism
for a variety of
spatiotemporal phenomena seen in real systems.

In this paper, we have analyzed in detail a model of generic relaxation
oscillators (each comprising activator and inactivating components) 
coupled to their nearest neighbors through lateral inhibition 
via diffusion of the inactivating component.
%species of the chemical reaction. 
The model reproduces all the spatiotemporal patterns
observed in the experiments mentioned earlier and its simplicity
allows an analytical understanding of their genesis.
In particular, we have given possibly the simplest theoretical
demonstration of the existence and stability of an anti-phase
synchronized state for coupled relaxation oscillators.
In addition to the patterns reported in experiments, we also observe
other states
such as, attractors corresponding to spatially co-existing dynamically distinct
configurations which we term chimera states. 
Although homogeneous arrays of generic relaxation oscillators have
been studied extensively, the observation of such spatially 
heterogeneous attractors for these systems is a novel finding of our
paper. 
%The robustness of these chimera states is demonstrated by a
%systematic characterization of the basins of attraction for
%various patterns seen in the model. 
We have performed for the first time a systematic characterization of
the basins of attraction for 
various patterns seen in the model that also demonstrates the
unexpected robustness of the chimera states
%The appreciable size of
%basins and absence of sensitive dependence on parameter values
and suggest that all the observed states can be obtained in suitable
experiments.
In addition, we report the occurrence of phase defect-like
discontinuities moving ballistically through the system that on
collision with each other can produce complex patterns. We
observe analogous structures in two-dimensional media that
have a striking resemblance to persistent
configurations in cellular automata (CA), e.g., ``gliders'' in the
``Game of Life'' CA~\cite{Conway}, which have been linked to 
the universal computation capabilities of such
systems~\cite{Wolfram,Cook04}. 
The observation of these patterns in our model system is remarkable 
considering the simplicity of the underlying dynamics.

\begin{figure}
\begin{center}
\includegraphics[width=0.99\linewidth]{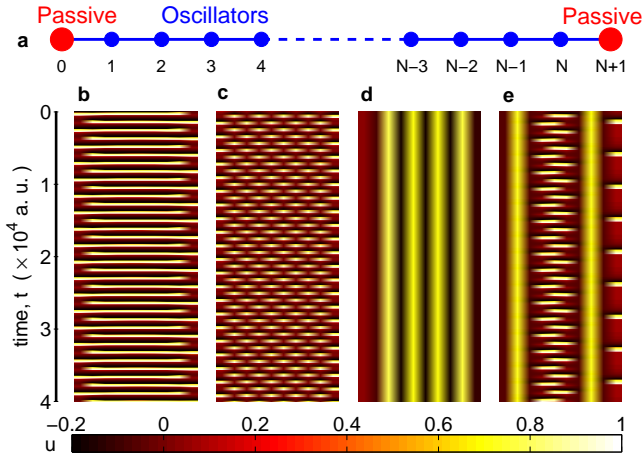}
\end{center}
\caption{(color online). Spatio-temporal evolution of
%model system 
a 1-dimensional array of coupled relaxation oscillators ($N=10$) 
with passive
elements at the boundaries [model system shown 
schematically in (a)]. Pseudocolor plots of the activation
variable $u$ indicate different regimes characterized by 
(b) synchronized oscillations (SO), (c) anti-phase synchronization
(APS), (d) 
spatially patterned oscillation death (SPOD) and (e) chimera state (CS), i.e.,
co-occurrence of spatial patches with dynamically distinct behavior.} 
\label{fig1}
\end{figure}

%Methods
The system we have analyzed comprises $N$ relaxation oscillators
interacting with each other in a specific topology.
For the dynamics of individual relaxation oscillators we have used the 
phenomenological FitzHugh-Nagumo (FHN) equations, which is a generic
model for such systems. Each oscillator is described by
a fast activation variable $u$ and a slow inactivation variable
$v$:
\begin{equation}
\begin{split}
\dot{u} &= f(u, v) = u\ (1 -u)\ (u -\alpha) - v,\\
\dot{v} &= g(u, v) = \epsilon\ (k\ u -v -b),
\end{split}
\end{equation}
where, $\alpha=0.139$, $k = 0.6$ are parameters describing the
kinetics, $\epsilon = 0.001$ characterizes the recovery rate
of the medium and $b$ is a measure of the asymmetry of the oscillator
(measured by the ratio of the time spent by the oscillator at high and
low value branches of $u$). Parameter values have been chosen such that the
system is in oscillatory regime. Small variations in the
values do not affect our results qualitatively.
To investigate spatial patterns generated by interaction between the 
oscillators, they are arranged in a 1-dimensional chain
[Fig.~\ref{fig1}~(a)]. In actual chemical experiments, the beads
containing the reactive solution are suspended in a chemically inert
medium which allows passage of only the inhibitory chemical
species~\cite{Toiya08}.
In our model, the oscillators are diffusively coupled 
via the inactivation variable $v$.
The boundary conditions for the chain take into account the inert
medium by including non-reactive passive elements at each end that are
diffusively coupled to the neighboring oscillators.
The inert medium between the oscillators are not considered
explicitly, their volume being relatively
small compared to the reservoirs at the boundary.
We have verified that inclusion of intermediate non-reactive cells
diffusively coupling each pair of oscillators do not affect the fixed-point 
equilibria of the system or their stability,
subject to suitable rescaling of the diffusion constant.
The dynamics of the resulting system is described by
\begin{equation}
  \begin{split}
    \dot{v}_0        &= D_v\ (v_1-v_0), \\
    \dot{u}_i        &= f(u_i, v_i), \\
    \dot{v}_i        &= g(u_i, v_i) + D_v\ (v_{i-1}+v_{i+1}-2\ v_i), \\
    \dot{v}_{_{N+1}} &= D_v\ (v_{_N}-v_{_{N+1}}),
  \end{split}
\end{equation}
where $i = 1, 2, \ldots, N$ and the diffusion constant $D_v$
represents the strength of coupling between neighboring relaxation
oscillators through their inactivation variables.
For most results reported here $N = 10$ although we have used higher
values of $N$ upto $1000$ to verify that the qualitative results are
not sensitively dependent on system size. We have also verified that
the boundary conditions do not sensitively affect the
results by considering periodic boundaries and observing
patterns qualitatively identical to those reported here.
The dynamical equations are
solved using an adaptive Runge-Kutta scheme (e.g., rkf45~\cite{GSL}).
The behavior of the system for each set of parameter values $b$
and $D_v$ is analyzed over many ($10^4$) initial conditions, with
each oscillator having a random phase chosen from a uniform
distribution.

Fig.~\ref{fig1}~(b-e) shows a variety of asymptotic 
spatio-temporal patterns that are
observed in the model system: synchronized oscillations (SO) with all elements
(except those at the boundary) having the same phase (b),
anti-phase synchronization (APS) with neighboring elements in opposite
phase (c), Spatially Patterned Oscillator Death (SPOD) regime where
the oscillators are arrested in various time-invariant states (d) and
Chimera States (CS) where oscillating regions co-exist with
patches showing negligible temporal variation (e).
However, these do not exhaust the range of possible spatio-temporal
phenomena that are observed including propagating structures
that are discussed later. Note that, both APS and
SPOD states have been observed experimentally in chemical systems.
Although the latter have sometimes been referred to as ``Turing
patterns'' in the literature, SPOD is distinct as it is not obtained
through destabilization of a homogeneous equilibrium (Turing
instability) but through a process of oscillator death.
%Note that, the ``Turing patterns" reported in
%the experiments with coupled chemical oscillators are identical to the
%SPOD states mentioned above. They are distinct from the classic Turing
%patterns as unlike the latter, SPOD is not arrived at by
%destabilization of a homogeneous equilibrium but rather through
%a process of oscillator death (the uncoupled elements having
%qualitatively distinct dynamics - viz., fixed point and limit cycle -
%in the two cases).
%Experiments have also reported the occurrence of the APS states, even when
%the initial state is synchronized, suggesting a large basin of
%attraction for APS in certain parameter regions. 

\begin{figure}
\begin{center}
\includegraphics[width=0.99\linewidth]{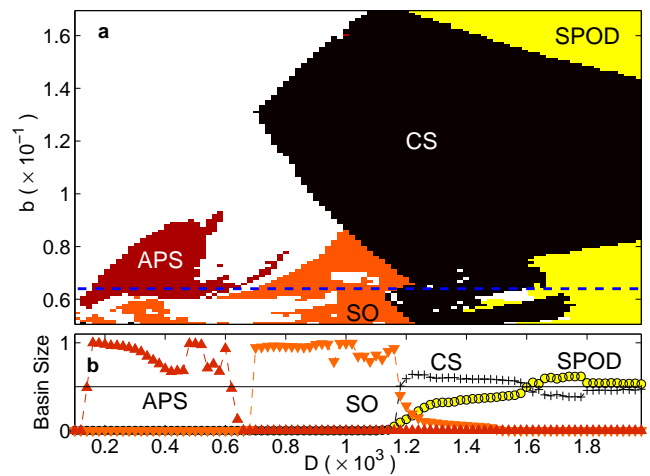}
\end{center}
\caption{(color online). Different dynamical regimes of a
1-dimensional array of coupled relaxation
oscillators ($N = 10$) in the $D_v-b$ parameter plane showing regions 
where the majority ($>50\%$) of initial conditions result in
synchronized oscillations (SO), anti-phase synchronization
(APS), spatially patterned oscillator death (SPOD) and chimera
state (CS). (b) Variation of the attraction basin size for the different
regimes mentioned above (measured as fraction of initial states
reaching the attractor) with coupling strength $D$ for $b=0.064$
[i.e., along the broken line shown in (a)].
In practice, the regimes are distinguished by thresholds applied on the
order parameters $\sigma^2_t(u_i)$, $\langle
\sigma^2_i(u) \rangle_t$, $\langle \sigma^2_{even}(u) \rangle_t$
and $\langle \sigma^2_{odd}(u) \rangle_t$, which have been taken to be
0.05 for the present figure. Basin sizes have been estimated using 
$10^4$ initial conditions.}
\label{fig2} 
\end{figure}

To investigate the robustness of the observed patterns
in detail, we numerically estimate the size of their basins of
attraction in the ($b, D_v$) parameter space
(Fig.~\ref{fig2}).
This provides an indication as to whether a pattern can be
observed in actual experiments as for this they must be immune to
the unavoidable noise present in any experiment.
%  DISCUSSION OF THE ORDER PARAMETERS
In order to segregate the distinct pattern regimes in the ($b,D_v$)
space [Fig.~\ref{fig2}~(a)]
we introduce the following order parameters.
The number of non-oscillating cells in the bulk of
the system, $N_{no}$, i.e., cells for which the variance of the
activation variable $u$, $\sigma^2_t(u_i)$, is zero, is used to characterize the SPOD
%variance of $u$ variable is less than a threshold ($0.05$).
($N_{no} = N$) and CS regimes ($0 < N_{no} < N$).
%2< N_{no} < N
%We have taken the boundary into account in that if less than three
%cells are non-oscillating we don't call it Chimera.
The SO and APS states both have all elements in the bulk oscillating.
However, SO is distinguished by having all oscillators in the same
phase as measured by the variance of 
the activation variables $u$, $\langle \sigma^2_i(u) \rangle_t$ = 0,
where $\langle ~\rangle_t$ represents time average. 
%The second order parameter is time-averaged variance of the $u$
%variable for all the oscillators : $\langle \sigma^2_i(u) \rangle_t$.
%This order parameter will be small ($<0.05$) for synchronized
%oscillations.
We can also define the synchronization among the oscillators in two
distinct sub-lattices, namely, those which occur at even-numbered and
those which occur at odd-numbered sites of the chain, as measured by
the time-averaged variance of the activation variable, viz.,
$\langle \sigma^2_{even}(u) \rangle_t$ and
$\langle \sigma^2_{odd}(u) \rangle_t$.
This pair of order parameters are zero for both SO and APS states; 
however, if $\langle \sigma^2_i(u) \rangle_t >0$, it signifies the APS
regime.
%system is not in the SO state then small values of the pair will imply
%APS state.
In practice, the different
regimes are characterized by thresholds whose specific values
do not affect the qualitative nature of the results.
Fig.~\ref{fig2}~(a) indicates the parameter regions where SO, APS, SPOD
and CS states are observed for 
more than $50\%$ of initial conditions (i.e., they have the
largest basin).
As already mentioned, the system also exhibits other regimes apart from
the above ones, which occur in
regions of ($b, D_v$) parameter space shown in white. 
%All the patterns shown in
%Fig.~1~(b-e) are therefore sufficiently robust to be seen in
%experiments performed under suitable conditions.

%Discussion of Fig2(b)
%Distinction between unstable patterns and pattern attractors with
%varying basin sizes
While diffusive coupling in a homogeneous system of oscillators is
expected to promote the SO state~\cite{Kurths}, 
a striking result from this phase diagram is that for certain
parameter values the APS state has a very large basin of attraction
[Fig.~\ref{fig2}(b)].
%so that almost all of the random initial conditions
%end up in the APS state.
The existence of APS 
%for diffusively coupled homogeneous system
is somewhat counter-intuitive as for diffusively
coupled identical isochronous oscillators the only stable attractors
are synchronized oscillations or oscillator death
\cite{Kurths}. %Torre,
While anti-phase synchronization has been seen earlier in a 
pair of identical oscillators~\cite{Rand80}, it is
not obvious that APS will have a large basin
of attraction for an array of oscillators.
%It is expected that SO will always have large basin whenever it
%exists 
To understand the origin of such anti-phase oscillations 
we consider a simple model that captures the essence of 
relaxation oscillation phenomena and can be solved exactly.
%
%Sometimes we can have an attractor in a dynamical system whose basin
%of attraction is so small that it is very difficult to see them in any
%real situation.
%The attractors which we see in real experiments must have a reasonably
%large basin of attraction so that they are robust against the
%unavoidable noise present in any experimental situation.
%To check the robustness of the various attractors, we estimate
%the volume of basin of attraction for each of the asymptotic behavior
%for different values of $b$ and the coupling $D$.
%
\begin{figure}[t]
\centering
\includegraphics[width=0.99\linewidth]{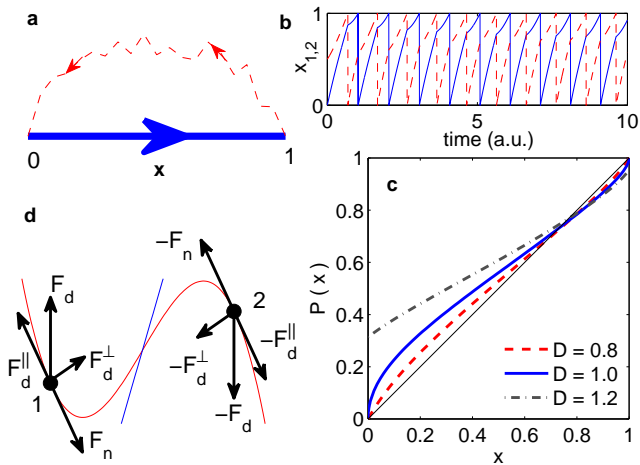}
\caption{(color online). (a) Schematic diagram of a limit cycle trajectory for an
oscillator in the
relaxation limit ($\epsilon \rightarrow 0$) and extreme asymmetry (for
details see text) such that the oscillator is on
the solid line ($0 < x < 1$) for its entire period. 
(b) Time-series of two such coupled oscillators
[Eqn.~(\ref{eqn3}) with $D=1$] and (c) the
Poincare map for the system at different coupling strengths $D$
showing stable anti-phase synchronization.
(d) Phase-plane diagram
%of an oscillator 
%with $v$-nullcline intersects
%the $u$-nullcline symmetrically, 
indicating the general mechanism (see text) for
oscillator death in a system of two coupled oscillators (1 and 2).} 
\label{fig3}
\end{figure}
%    ANALYTICAL SOLUTION OF A SIMPLE DYNAMICAL SYSTEM
%\section*{Analytical solution of a simple model showing anti-phase oscillation}
%We analyze a dynamical system which explains the origin of anti-phase
%oscillations by diffusive coupling.
%INETRPRETATION
%In the limit of extreme asymmetry (governed by b) 
%for the FHN oscillator system (Eq 1) such that 
%the time taken in the fast branch of the u nullcline is
%zero, the slow variable is the varable x of the new model
%As a relaxation oscillator can be generally seen as a dynamical system
%with segments with different speeds and our model is the extreme case
%in which one part is much slower than the rest of the limit cycle.
We consider the relaxation limit ($\epsilon \rightarrow 0$ in FHN
system) and extreme asymmetry where the limit cycle has a slow segment
in which the system spends the entire duration of the oscillation period (the
remaining segment of the cycle being traversed extremely fast).
Under these considerations, we obtain the one-dimensional dynamical
system:
$ \dot{x} = \omega(x),$
where $x$ parameterizes the slow part of the limit cycle and can be
redefined to belong to the interval $(0,1)$.
Fig.~\ref{fig3}~(a) shows a schematic diagram of the
trajectory of the limit cycle, where the system spends almost its
entire oscillation period on the solid branch (the return from $x=1$
to $x=0$, shown by the broken line, is considered to be instantaneous).
%alternatively be thought to be the fractional part of a linear
%dynamical system.
The model can be exactly solved if $\omega(x)$ is a constant ($= \omega$,
say), although the geometrical argument is valid for any arbitrary 
positive definite function defined over the interval $(0,1)$.
By appropriate choice of time scale, we can set the period
$\omega^{-1} = 1$ without loss of generality.
%\subsection*{Two Coupled Identical Systems}
A system of two such diffusively coupled oscillators can be described by 
\begin{equation}
  \dot{x}_{1} = 1 + D\ (x_{2}-x_{1}),\quad
  \dot{x}_{2} = 1 + D\ (x_{1}-x_{2}).
% \dot{x}_2 &= 1 + D\ (x_1-x_2)
\label{eqn3}
\end{equation}
Given the values of $x_1, x_2$ at some arbitrary initial time
$t'$,
the solution of Eqn.~(\ref{eqn3}) at a later time $t$ follow the relations:
%The linear combination of the solution of this linear system of
%equation before either oscillator reaches $x=1$ is given by
\begin{equation}
\begin{split}
 x_1(t) + x_2(t) &= x_1(t') + x_2(t') + 2 (t-t'), \\
 x_1(t) - x_2(t) &= [x_1(t') - x_2(t')]\ \exp[-2D(t-t')],
\end{split}
\label{eqn4}
\end{equation}
till time $t''$ when $max (x_1, x_2)$ reaches $x=1$. After this the
larger of $x_1,x_2$ is 
%where, at every instant $0 \leq x_1 (t), x_2 (t) \leq 1$.
mapped back to
$x=0$ (because of the instantaneous nature of the remaining segment of
the limit cycle) and $t'$ in Eqn.~(\ref{eqn4}) is replaced by $t''$.
%\subsection*{Poincare Map}
Using the above exact solution of the coupled system ($\ref{eqn3}$),
its Poincare map $P(x)$ is constructed in two steps. First, 
if $x_1$ starts at $0$ and $x_2$ starts at some point $0 < x < 1$, 
we find the location of $x_1[=f(x)]$ at some time $t$ 
when $x_2=1$ (which is then
immediately mapped to $x_2=0$). Now, starting with $x_2 = 0$ and
$x_1=f(x)$, when $x_1=1$ we find the location of $x_2$: $x'=
f(f(x)) = P(x)$.
Using Eqn.~(\ref{eqn4}), with $x_1(t')=0$, $x_2 (t')=x$, $x_1 (t)=f(x)$
and $x_2(t)=1$, we get 
%Let at time $t$, $x_2$ reaches $1$, i.e.
%$$ x_1(0) = 0, x_2(0) = x, x_1(t) = f(x), x_2(t) = 1.$$
%Using the exact solution, we get -
%\begin{equation}
%\begin{split}
%    t &= 1 - x / 2 + (2D)^{-1}\ W[-Dx\ \exp\{D(x-2)\}] \\
% f(x) &= 1 + D^{-1}\ W[-Dx\ \exp\{D(x-2)\}]
%\end{split}
%\end{equation}
\begin{equation}
f(x) = 1 + D^{-1}\ W[-Dx\ \exp\{D(x-2)\}],
\label{eqn5}
\end{equation}
where $W$ is the Lambert W-function.
Fig.~\ref{fig3}~(b) shows the Poincare map $P(x)=f(f(x))$ 
for different values of the
coupling strength $D$. The map has one stable and one unstable fixed
point, which correspond to the anti-phase synchronized (APS) and
synchronized oscillating (SO) states, respectively.
Thus, for the model (\ref{eqn3})
we find that APS state not only exists but is the only stable state.
Relaxing the extremal conditions under which this was derived may 
allow a stable SO state to coexist with the stable APS
state~\cite{Izhikevich_2000}. The derivation we have shown here
is possibly the simplest one exposing
the fundamental mechanism for generating APS states in
any system of diffusively coupled oscillators that can exhibit
anti-phase oscillations.

%Why SPOD ? 
When the coupling $D_v$ between oscillators in the array is increased
to very high values, we observe that the oscillatory regimes (e.g.,
SO and APS) are replaced by time-invariant spatial patterns such as
SPOD (Fig.~\ref{fig2}). To understand the genesis of SPOD at strong
coupling, we can again focus on a pair of coupled relaxation
oscillators in the relaxation limit
($\epsilon \rightarrow 0$). The parameter $b$ is chosen such that 
the $v$-nullcline is placed
symmetrically between the two branches of the
$u$-nullcline 
with the oscillator taking equal time to traverse each branch
[Fig.~\ref{fig3}~(d)].
%on which the oscillator spends the entire duration of
%its period [Fig.~\ref{fig3}~(d)].
When the two oscillators (1 and 2) are in opposite branches (as shown
in the schematic diagram), the two opposing forces acting on each
oscillator, corresponding to the coupling [$F_d = D_v (v_2-v_1)$] and
the
intrinsic kinetics ($F_n$)
respectively, can exactly cancel when the coupling is strong resulting
in oscillator death.
%When the coupling between the two oscillators (1 and 2) is strong,
%if the oscillator states are in appropriate
%positions in the opposite branches 
%(as shown in the schematic diagram), the two opposing forces acting on each
%oscillator, corresponding to the coupling [$F_d = D_v (v_2-v_1)$] and the
%intrinsic kinetics ($F_n$)
%respectively, exactly cancel and gives rise to a new fixed point 
%for the system. 
Note that only the component of $F_d$ parallel to the
intrinsic force $F_n$ ($F_d^{||}$) needs to balance $F_n$ as
$F_d^{\perp}$ has no effect in the relaxation limit.
The symmetry of the oscillator ensures that the force due to the
intrinsic kinetics ($F_n$) for the two oscillators
are identical but oppositely directed in the steady state.
%This results in all forces being exactly opposite for the two
%oscillators; in absence of symmetry, this is no longer the case.
%only F_d would be same but not F_n and hence even the components of
%F_d would be different.
The occurrence and stabilization of this {\em heterogeneous} 
time-invariant state (as
1 and 2 need to be in different branches, corresponding to very
different values of $u_{1,2}$)
is the key to the occurrence of SPOD at strong coupling.
%which
%is quite distinct from the classic Turing pattern formation
%scenario (as mentioned above).
At intermediate values of coupling $D_v$ between oscillators in a
large array, the
competition of this mechanism with the natural oscillatory dynamics
(seen at low coupling) may give rise to chimera states that exhibit
characteristics of both, i.e., containing patches of elements that are
oscillating as well as segments that show SPOD-like pattern.
%      DISCUSSION OF TURING-LIKE (SPOD) AND CHIMERA PATTERNS
%      SPOD : Spatially Patterned Oscillator Death
%Returning to the array of oscillating elements, we observe that when
%the coupling $D$ becomes very strong, the different oscillatory regimes
%(such as SO and APS) give way to attractors corresponding to
%time-invariant spatial patterns (SPOD). 
%In between these dynamically
%very different behavior, we obtain chimera states that exhibit
%characteristics of both, i.e., containing patches of elements that are
%oscillating as well as segments that show SPOD-like pattern.
This CS regime is especially interesting as the system exhibits a
strikingly heterogeneous dynamical state in spite of the bulk of the
array being homogeneous with identical elements coupled in the
same fashion. We have explicitly verified
that the occurrence of CS is not dependent on boundary conditions
by reproducing it also in systems with periodic boundaries.
The observation of such states in a generic model of relaxation
oscillators suggests that they should be present in a wide class of
systems, and indeed similar phenomena have been recently reported in 
a specific chemical system model~\cite{Vanag11}.
Note that, the chimera state described here comprises regions with 
dynamically distinct behavior, and is different from its namesake that
refers to co-occurrence of coherent and
non-coherent domains~\cite{Strogatz2004}. 
%Is chimera a result of competition between oscillation and
%oscillation death mechanisms
%The reason why people may not have reported CS even though they may
%have observed it in experiments is because in the latter they can be
%ascribed to the inherent heterogeneity of experimental systems -
%whereas our system by design is homogeneous in the bulk

%  FIGURE 4
\begin{figure}[t]
\centering
\includegraphics[width=0.99\linewidth]{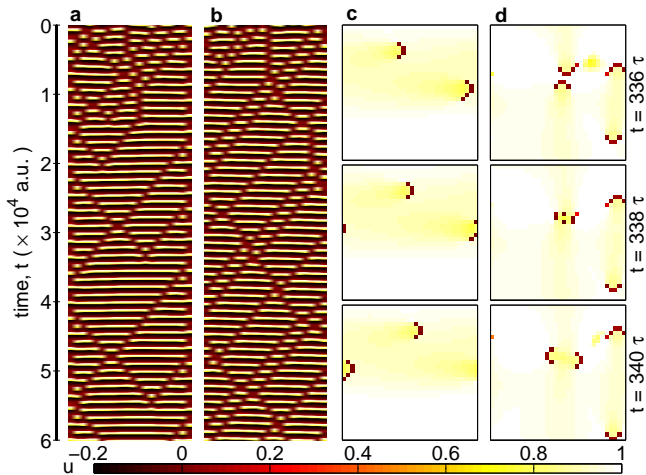}
\caption{(color online). (a-b) Spatio-temporal evolution of a system of coupled
relaxation oscillators showing traveling waves of phase defects in (a)
a linear array with passive elements at the boundaries and (b) with
periodic boundary conditions. (c-d) Propagating defects in 
two-dimensional media with periodic boundary condition showing (c) two
horizontally moving ``gliders" and (d) collision of two ``gliders".
For clear visualization of the motion of the spatially extended defects,
snapshots of the two-dimensional medium are taken at intervals which
are multiples of
the oscillation period for the mean activity of the system $\tau$.
Animations are available in Ref.~\cite{supp}.
%Stroboscopic snapshots of the time-evolution of the system make the
%defects,
%which are now spatially extended objects, clearly visible.
} 
\label{fig4}
\end{figure}
Apart from the spatio-temporal patterns shown in
Fig.~\ref{fig1}~(b-e) we also observe attractors
corresponding to point-like ``phase defects'' (i.e., there is a
discontinuity of phase along the oscillator array at this point)
moving in the background of
system-wide oscillations.
As seen from a typical example of such states [Fig.~\ref{fig4}~(a)], 
after initial transients these defects move in the
medium with interactions
between two such entities resulting in either the two getting
deflected in opposite directions, or
either one or both getting annihilated (unlike defects in
non-oscillatory media such as domain walls which annihilate on
collision~\cite{ChaikinLubensky}).
While the boundary for systems with passive elements at the ends is a
source of new defects entering the medium, similar persistent
structures are also seen in systems with periodic boundary
conditions where, in the steady state,
a conserved number of defects can
reflect off each other indefinitely [Fig.~\ref{fig4}~(b)].

To observe how these propagating defects manifest in higher
dimensional systems, we consider a 2-dimensional array of coupled
oscillators defined on a torus~[Fig.~\ref{fig4}~(c-d)].
The system can have extremely complicated transient phenomena, and
for simplicity we show here only its asymptotic behavior.
For a square lattice, we observe that there is a specific
configuration of four sites that persists indefinitely
(reminiscent of the {\em glider} configurations in the 2-dimensional
CA ``Game of Life"~\cite{Conway})
and can move in horizontal or vertical directions.
%For a sufficiently large system size, a number of ``gliders'' 
%can be seen moving
%in different directions without interacting with each other.
Interaction of such ``gliders'' can produce complex spatio-temporal 
patterns. Fig.~\ref{fig4}~(d) shows two ``gliders'' that on collision move
off in directions perpendicular to their incident ones~\cite{supp}.
%These patterns remind us of 2-d cellular automata e.g. Conway's Game
%of Life.

%DISCUSSION
%These spatio-temporal evolutions can be used to perform computation
%as in CA
To conclude, we have shown that a simple model of relaxation
oscillators coupled via lateral inhibition can exhibit a wide variety
of striking spatio-temporal patterns. 
Our comprehensive investigation has revealed the global features of 
the dynamics and robustness of the attractors.
%By focusing on the basins of
%attraction of the different states, unlike most previous studies 
%which only looked at a few initial conditions, we have revealed the
%global features and robustness of these attractors.
%Our results not only explain the
%occurrence of several patterns previously observed in experiments on
%chemical systems, but moreover predict the occurrence of new ones that
%can be realized in different experimental conditions. 
As our results are based on a very generic model, it suggests 
that the patterns may
be observed in a range of experimental realizations, such as
electronic circuits implementing relaxation
oscillators~\cite{Scholl2010} and Pt wire undergoing CO oxidation where the
system is in an oscillatory regime~\cite{Ertl}, apart from
microfluidic chemical systems mentioned earlier.
Our initial exploration of propagating configurations in
2-dimensional media suggests that systems of higher dimensions 
may yield even more striking discoveries.
It is intriguing to explore the possibility of using the propagating
defects for computation~\cite{Liang1995,Epstein2007}, 
as analogous entities have been used to construct logic gates in 
CA~\cite{Conway}. This may tie
together two of the groundbreaking ideas of Alan Turing, 
%(whose birth centenary is being celebrated this year), 
viz., pattern formation through reaction-diffusion mechanism and universal
computation~\cite{Turing}.
%Most previous studies have focused on single attractors and have 
%not considered the basins of attraction of the
%different attractors - typically they
%have only looked at a few initial conditions and thereby missed out
%characterizing the global
%nature of the attractor dynamics.

We thank Bulbul Chakraborty and Sudeshna Sinha for helpful
discussions and the HPC facility at IMSc for providing
computational resources.

%{2}

\end{document}